*Technical Article*

# Aspects of Coupled Problems in Computational Electromagnetics Formulations

**Abstract** – This paper addresses different aspects of "coupled" model descriptions in computational electromagnetics. This includes domain decomposition, multiscale problems, multiple or hybrid discrete field formulation and multi-physics problems. Theoretical issues of accuracy, stability and numerical efficiency of the resulting formulations are addressed along with advantages and disadvantages of the various approaches. Examples for multi-method, multi-domain, multi-formulation and multi-physics coupled formulations stem from numerical testing of electromagnetic compatibility in complex scenarios and numerical dosimetry of biological organisms in electromagnetic exposure situations and from simulations of large systems in electromagnetic power transmission.

## I. INTRODUCTION

In the computation of electromagnetic fields many application problems involve the aspect that different model descriptions need to be coupled. This may involve the coupling of different computational domains as e.g. in domain decomposition or the combination of different discretization schemes to describe a problem. Problems combining relevant effects originating at very different scales also require coupled simulation approaches as those that involve physical models beyond the Maxwell regime (multi-physics). The challenge arising with these "multi-x" approaches lies in the accuracy, stability and numerical efficiency of the resulting coupled field formulations. While these aspects have been addressed in the past mostly related to the simulation problem at hand, more recent research efforts try to mathematically analyze these aspects in a more general way. In the following section a classification of coupled formulations in computational electromagnetics is proposed and common interfaces are discussed. Section 3 summarizes a general framework for coupled formulations based on dynamic iteration and discusses its convergence briefly.

This paper also presents some real-world applications stemming from electrical engineering in Section 4. Examples for coupled formulations involve e.g. the numerical testing of electromagnetic compatibility (EMC) in complex scenarios involving coupled field-cable formulations and numerical dosimetry of biological organisms in electromagnetic exposure situations, in order to ensure occupational safety and health. Also of interest is the simulation of complex systems in electromagnetic power transmission, e.g. the simulation of high voltage insulators, cable terminators or energy cables, which often require coupled electromagnetic field and thermal simulations. These formulations may feature different methods (e.g. discretization methods) that are employed on different subdomains. Often, exploiting this degree of freedom allows more efficient simulations for problems, e.g. coupling of Finite Elements and Boundary Elements Methods (FEM-BEM) for electro-quasistatic field simulations or using the Method of Moments (MoM) coupled to the Finite Difference Time Domain method (FDTD) [35] or Finite Integration Time Domain method (FITD) [33] for radio-frequency problems or e.g. magneto-quasistatic magnetic fields coupled with electric power networks [11].

## II. CLASSIFICATION

Many classical problems in electrical engineering are still simulated with sufficient accuracy by electromagnetic models based on Maxwell's equations without extension. On the other hand, as discussed before, some applications require multiphysical models. Over the last years, trends like miniaturization and elevation of frequencies increased multiscale and multirate phenomena. As a consequence, many applications, where traditionally a single model was sufficient, must now be represented by more complex descriptions. This results in an increased demand to consider coupled problems. The important cases belong to one or more of the following classes:

- Hybrid formulations, describing subdomains by different field quantities, e.g. current and magnetic vector potential based formulations
- Hybrid discretizations, e.g. using the finite element method for complicated geometries and boundary elements for uniform subdomains, e.g. [29]
- Coarsened and refined modeling, e.g. refining Maxwell's equations in a subdomain by a model that accounts for quantum effects or coarsening the problem by describing subdomains by electric circuits, i.e., disregarding spatial distributions, [11,26]
- Multiphysics, e.g. coupling electromagnetic field problems with motion effects, thermodynamics and fluid dynamics

Sometimes the additional effort due to coupling is negligible and only small modifications are necessary. This is often the case for refined modeling, e.g. thin sheets are nowadays incorporated into standard finite elements codes. The same holds true for simple multiphysical examples where some effects can be extracted in a post-processing step because their influence is negligible, i.e., the coupling is only one-way, see Fig. 1a). In practice, the more challenging multiphysical problems require the solution of an additional mutually coupled transient process that is described by its own set of (partial-) differential equations. The interfaces are typically given by

- prescribing boundary conditions, i.e., a part of the degrees of freedom is fixed by another problem (e.g. application of a voltage drop as Dirichlet condition)
- imposing a source, i.e., introducing an external contribution with respect to another problem without removing degrees of freedom (e.g. imposing a source current from a circuit)

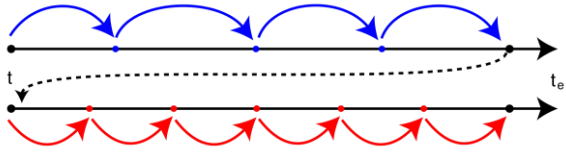
One-way Coupling / Parameter Extraction

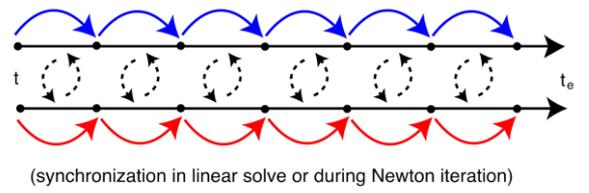
"Strong" Coupling

(synchronization in linear solve or during Newton iteration)

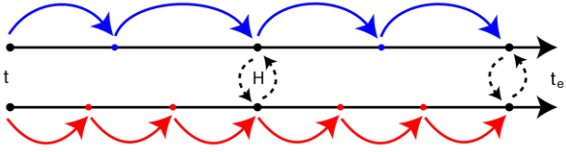
"Weak" Coupling

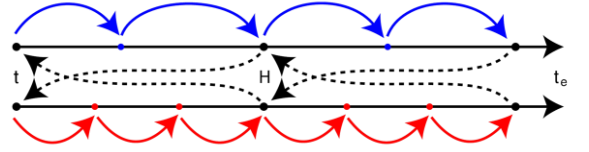
Dynamic Iteration / Waveform Relaxation

**Figure 1:** Coupling schemes: (a) one-way coupling, (b) strong coupling, (c) weak coupling and (d) dynamic iteration

- modification of material parameters, i.e., changing the value with respect to degrees of freedom of another problem (e.g. a conductivity depends on the temperature described by a heat equation)
- deforming the computational domain, i.e., the domain is decomposed and the subdomains move with respect to another problem (e.g. the rotor position of an electrical machine is described by a motion equation)

In either case, the scaling of the quantities becomes increasingly difficult, often the problems are worse conditioned and properties of the equations that could be exploited before such as sparsity patterns, symmetry and definiteness may be lost. Furthermore, the usage of black-box solvers does rarely allow for a strong coupled formulation. Thus it is often beneficial to solve each subproblem separately instead of monolithically (i.e., strongly coupled), see Fig. 1.

### III. CONVERGENCE OF COUPLED SIMULATIONS

When having mutual dependencies and solving subproblems independently, there is commonly a lack of information ("splitting error"). Hence it is important to ensure a sufficient data exchange. Often the problems are only weakly coupled in time. Then a small time lag is acceptable and the problems are synchronized only at discrete time points, Fig. 1c). This approach is not stable if the interaction of the subsystems becomes stronger. It might not even converge when the frequency of the data exchanged is increased. This problem can be overcome by an iterative procedure. In the following we discuss the dynamic iteration scheme that is proven to be convergent for many applications.

Let us consider coupled problems in the time domain following the method of lines, i.e., space is discretized first, for example using FEM. The equations are semi-discrete and depend only on time. For simplicity of notation only semi-explicit systems are considered of the form

$$\frac{d}{dt}\mathbf{y} = \mathbf{f}(\mathbf{y},\mathbf{z}) \quad \text{and} \quad \mathbf{0} = \mathbf{g}(\mathbf{y},\mathbf{z})$$

on the time window $t \in [0,H]$ with initial values $\mathbf{y}_0$ and $\mathbf{z}_0$ where the Jacobian of $\mathbf{g}$ with respect to $\mathbf{z}$ is regular $\det(\partial \mathbf{g}/\partial \mathbf{z}) \neq 0$. This formulation is a first order differential-algebraic equation of index 1. It is easily shown that many implicit formulations, e.g., the magneto- and electro-quasistatic field equations, heat phenomena and many other multiphysical problems (see Section 4) can by represented equivalently in this form. Left-multiplication of the system by the inverse (or pseudoinverse) of the mass matrix, i.e.,

$$\mathbf{M}\frac{d}{dt}\mathbf{x} = \mathbf{F}(\mathbf{x}) \quad \text{multiplied by} \quad \mathbf{M}^+$$

results in a semi-explicit system as stated above. Here, the number of algebraic constraints depends on the dimension of the kernel of $\mathbf{M}$, for example the mass matrix in the curl-curl equation for magneto-quasistatic fields has a large kernel, due to non-conducting regions, while the classical heat equation has no algebraic constraints.

Time dependency in the right-hand-side function $\mathbf{f}$ and equations with higher order time derivatives can also be converted into the form above by introducing additional unknowns, e.g.,

$$\mathbf{M}\frac{d^2}{dt^2}\mathbf{x} = \mathbf{f}(t,\mathbf{x}) \Leftrightarrow \frac{d}{dt}\mathbf{x} = \mathbf{w} \quad \text{and} \quad \mathbf{M}\frac{d}{dt}\mathbf{w} = \mathbf{f}(t,\mathbf{x})$$

Please note, that these transformations are convenient for discussing the convergence property in a dynamic iteration, because it splits the degrees of freedom into variables that are defined by differential and algebraic equations. These variables exhibit different numerical behavior. In practice, the required operations are usually not necessary and probably impossible due to their computational complexity.

Using the convenient notation from above, the iterative mutual coupling of several problems yields the following set of equations (a Gauss-Seidel-type scheme)

$$\frac{d}{dt}\mathbf{y}_1^{(k+1)} = \mathbf{f}_1(\mathbf{y}_1^{(k+1)},\mathbf{z}_1^{(k+1)},\mathbf{y}_2^{(k)},\mathbf{z}_2^{(k)},...,\mathbf{y}_n^{(k)},\mathbf{z}_n^{(k)})$$

$$0 = \mathbf{g}_1(\mathbf{y}_1^{(k+1)},\mathbf{z}_1^{(k+1)},\mathbf{y}_2^{(k)},\mathbf{z}_2^{(k)},...,\mathbf{y}_n^{(k)},\mathbf{z}_n^{(k)})$$

$$\vdots$$

$$\frac{d}{dt}\mathbf{y}_n^{(k+1)} = \mathbf{f}_n(\mathbf{y}_1^{(k+1)},\mathbf{z}_1^{(k+1)},\mathbf{y}_2^{(k+1)},\mathbf{z}_2^{(k+1)},...,\mathbf{y}_n^{(k+1)},\mathbf{z}_n^{(k+1)})$$

$$0 = \mathbf{g}_n(\mathbf{y}_1^{(k+1)},\mathbf{z}_1^{(k+1)},\mathbf{y}_2^{(k+1)},\mathbf{z}_2^{(k+1)},...,\mathbf{y}_n^{(k+1)},\mathbf{z}_n^{(k+1)})$$

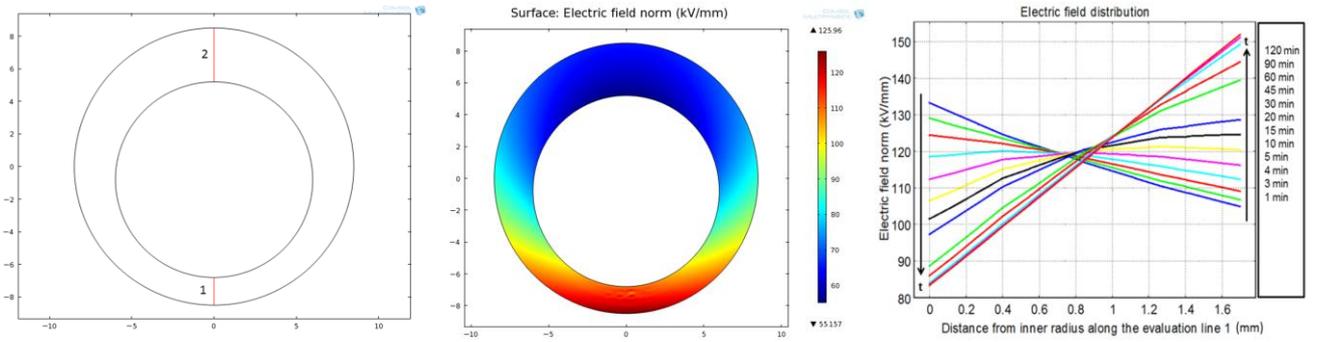

**Figure 2:** (a) Rotationally non-symmetric model and 2 evaluation lines in red, (b) electric field distribution during heating process and (c) electric field distribution along the evaluation line 1 in (a) during the heating process [34]

where $\mathbf{y}_j^{(k)}(t)$ and $\mathbf{z}_j^{(k)}(t)$ denote the solution of the $j$-th subsystem ($j=1,...,n$) after $k$ iterations ($k=1,...,m$) on the time window $t \in [0,H]$. In practice these solutions are obtained sequentially by a time-stepping scheme with a sufficiently high accuracy. Having solved one time window, i.e., the splitting error has been reduced below a certain tolerance, one proceeds to the next one. In [2,5] it has been shown that the scheme above corresponds to a fixed point iteration in function space

$$\begin{pmatrix} \delta_y^{(k+1)} \\ \delta_z^{(k+1)} \end{pmatrix} \leq \begin{pmatrix} O(H) & O(H) \\ \beta + O(H) & \alpha + O(H) \end{pmatrix} \begin{pmatrix} \delta_y^{(k)} \\ \delta_z^{(k)} \end{pmatrix}$$

with splitting errors $\delta_y^{(k)}$ and $\delta_z^{(k)}$ for the differential and algebraic components after $k$ dynamic iterations. The scheme above converges for sufficiently small time windows H if the contraction factor $\alpha$ is small enough. This corresponds to a weak coupling between old and new algebraic variables:

$$\left\| \left(\frac{\partial \mathbf{g}}{\partial \mathbf{z}^{(k+1)}}\right)^{-1} \frac{\partial \mathbf{g}}{\partial \mathbf{z}^{(k)}} \right\| < 1.$$

Then the spectral radius of the fixed point iteration operator can be made arbitrarily small and this guarantees convergence. In [2,5,26] also the overall stability with respect to the propagation of splitting errors on multiple windows has been analyzed.

In many relevant applications the systems $1,...,n$ can be reordered in such a way that there is no coupling between old and new algebraic variables and thus the contraction factor $\alpha$ vanishes completely. In particular, ordinary differential equations do not suffer from a divergent iteration because an algebraic coupling as discussed above cannot occur.

A vanishing contraction factor $\alpha$ guarantees a convergence rate of at least $O(H)$ such that only a few iterations are necessary to obtain a splitting error that is in the order of the discretization error of the time stepping scheme. Recent analysis shows that convergence rates up to order $O(H^2)$ can be achieved when choosing the coupling interface carefully.

The results above apply in particular to low frequent field-circuit, semiconductor-circuit, and field-thermodynamic coupled application. For example weak electromagnetic-thermal and FDTD-MoM couplings are discussed in the following sections.

## IV. COUPLED FORMULATIONS IN CEM

In this chapter some typical examples of coupled problems in computational electromagnetics are addressed that showcase the different coupling possibilities and techniques involved.

### A. Multiple Physical Models

From the large number of multi-physically coupled formulations in electromagnetic field simulations especially the coupling to thermal effects is widely established. As Joule heat losses typically are the main parasitic loss mechanism for many electromagnetic field problems with an impact on electromagnetic material parameters. This usually requires the additional simulation of thermostatic problems or even thermodynamic problems involving the solution of the heat transport equation. Since the characteristic time of a thermal process often exceeds that of the coupled electromagnetic process, usually a weak coupling approach can be adopted, cf. Fig. 1c). Here, the thermal process is the dominant process, which requires the Joule losses as thermal field excitation source to be calculated first in a post-processing step to an electromagnetic field problem solution. In turn, the electromagnetic material parameters may depend on the temperature and thus an iterative coupling as discussed in Section 3 might by necessary, Fig. 1d).

An example of such a coupled electrodynamic-thermodynamic problem is the simulation of a high-voltage direct current (HVDC) cable, where the dielectric PE (Polyethylene) cable insulator material features a low nonlinear conductivity

$$k(T, |\vec{E}|) = k_0 \cdot \exp(\alpha \cdot T) \cdot \exp(\beta \cdot |\vec{E}|),$$

where $k_0$ is a specific direct current conductivity at 0℃ and 0 kV/mm, $T$ is the temperature in ℃, $|\vec{E}|$ the value of electric field in kV/mm, $\alpha$ a temperature coefficient of the specific direct current conductivity and $\beta$ is a field strength coefficient of the specific direct current conductivity. The typical values for Polyethylene are $\alpha \sim 0.1\text{K}^{-1}$ and $\beta \sim 0.1\text{mm/kV}$.

Extending the one-dimensional coupled field formulation in [3], a coupled 2D simulation of these electric field energy transport systems [34] features the weakly coupled solution of a stationary current problem solving

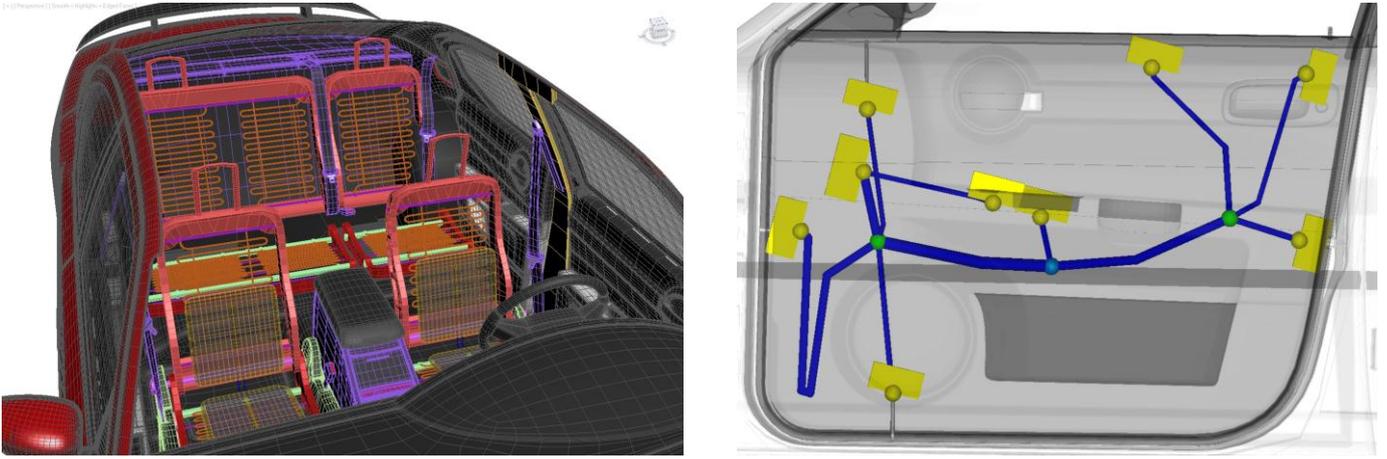

**Figure 3**: Car model including door harness model with blue cable bundles and yellow connectors [23]

$$\operatorname{div}(k(T,\varphi) \cdot \operatorname{grad}\varphi) = 0$$

where $T$ is the temperature in $K$, $\varphi$ is the electric scalar potential and $k(T,\varphi)$ is the nonlinear electric conductivity, and the transient heat transport equation

$$C_p \cdot \rho \cdot \partial_t T - \operatorname{div}(\lambda \cdot \operatorname{grad} T) = Q_E,$$

where $C_p$ is the heat capacity at constant pressure, $\rho$ is the density, $\lambda$ is the thermal conductivity, $Q_E$ describes electric heat sources calculated from the Joule heating from the cable and the thermal power losses from in the PE insulation material.

These simulations yield the (contra-intuitive) result that electric field stresses may be larger at the outer radius of such a cable than close to the inner conductor. Fig. 2 shows the model of a radially non-symmetric configuration: Due to high temperatures, the insulation thermoplastic material, e.g. LDPE (low density Polyethylene), may become soft and thus the inner conductor may sag because of gravity [21].

Beyond the coupled simulation of thermal and electromagnetic fields, other multi-physical models include e.g. acoustic simulations of electric machine noise emissions or electro-mechanical simulations [17]. Also the simulation of charged particles with so-called particle-in-cell codes, combining the solution of Maxwell's equations e.g. via FDTD or DGFEM schemes and Vlassov charged particle dynamics is a widely established standard specifically in accelerator design, see. e.g. [25].

### B. Refined Modeling

Different discretization methods within one electromagnetic field phenomenon are typically applied if one method alone is not capable of representing all the required features of an electromagnetic problem and/or if the benefit of the additional modeling capabilities gained outweighs any additional difficulties arising from the coupling process.

For radio-frequency electromagnetic wave propagation problems typically implementations of volume oriented space and time discretization methods such as the FDTD/FITD method or more recently Discontinuous Galerkin Finite-Element (DGFEM) methods are considered established standard simulation tools. They allow for a high resolution of geometric features in the near field of electromagnetic radiation sources. However, many electromagnetic compatibility (EMC) problems feature geometric aspect ratios which would require very large numbers of volume grid cells to cover all the relevant geometric details. This also affects the time step size of explicit methods due to the Courant-Friedrich-Levy (CFL) condition. While considerable progress has been made in terms of reducing computation times to an acceptable level even for large scale problems featuring hundred million grid cells by using graphic process units (GPU) accelerators, many EMC problems featuring both complex dielectric material distributions and large aspect ratios can still not be tackled with such a one-method approach.

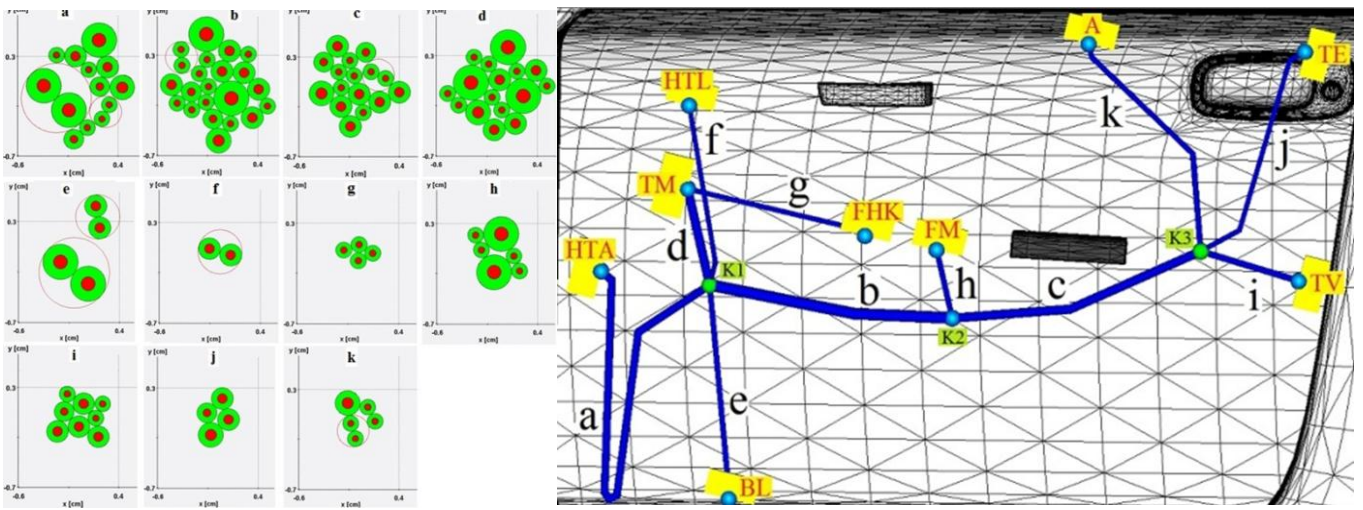

**Figure 4**: Door harness with yellow connectors, blue cable bundles and a cross section overview [23]

A typical example of this problem type arises from testing cable harness systems in complex electromagnetic environments, i.e., metallic enclosures (possibly with apertures) such as car bodies, airplane hulls or electronic system enclosures, see e.g. Fig. 3. As it is not possible to discretize the cable harness, where cables have a diameter in the (sub-) millimeter range, as part of a resonant structure with a typical size in the order of decameters or meters, a multi-method approach should be adopted. It combines a multiconductor transmission line (MTL) approach for the harness and a standard volume oriented discretization scheme for the surrounding field environment.

As an example for this approach, a complex harness system inside a car door (Fig. 4) is considered using a co-simulation approach combining two dedicated implementations available e.g. for the MTL model in the CST Cable Studio connected to a FITD implementation of the CST MicroWaveStudio [10]. Starting from a real door harness supplied by an industrial partner, the cable routes and their connector places are studied inside a high resolution 3D vehicle model [23].

The door harness consists of 32 cable routes with single wire and twisted pair cables. Within the MTL model an equivalent lumped element circuit is calculated with respect to the cable characteristics as e.g. 2D cross sections, cable lengths, Ohmic and dielectric losses and simulation frequency. Using static simulations in the 2D cross sections of the harnesses, the MTL inductance, capacitance and conductance matrices are computed. In addition, every single cable end, represented by a pin in the equivalent circuit, can be equipped with further lumped elements as e.g. a voltage source.

The MTL harness model is coupled via voltages and currents to the FITD model. The coupling can be performed by a uni-directional method (one-way coupling): For simulating a field radiation into the cable harness the electric field components of the FITD grid edges are integrated along the cable segments and each cable segment is then exposed to the resulting voltage (field-to-cable coupling). Alternatively, the currents calculated from the MTL model are used as field irradiation source into a reference conductor of the FITD model. Also, a bi-directional coupling (e.g. using dynamic iteration) is possible combining the simulation radiation and irradiation simultaneously, e.g. for resonating structures, by exchanging voltage and current information between the 3D solver and MTL harness model to calculate the fields.

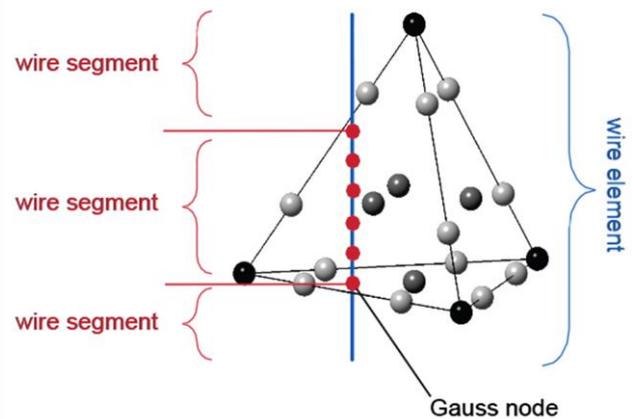

**Figure 5:** Wire-to-tetrahedron coupling using Gauss points along a wire segment cutting the tetrahedron [16]

For both uni- and bi-directional field coupling methods, the currents of the cable bundles are summed up to a single common mode current which is represented by an impressed field source in the 3D field model, while the cables are not physically present during the FITD calculation [10]. This approach often allows for a sufficient approximation of common mode currents in cable harnesses which pose a serious EMC problem in many technical systems.

Alternatively, field-cable coupled formulations are also possible using a strong coupling e.g. for full transient simulations. An approach used by Holland [20] to couple the FDTD scheme to a cable transmission line mode was recently extended to the higher order DGFEM formulations in [15]. Here, Maxwell equations discretized with a higher order DGFEM nodal formulation [19] are coupled to the cable transmission line model which is also discretized using the DGFEM approach [16]. For each finite element the Maxwell curl equations are reformulated to a system of ordinary differential equations

$$\frac{d}{dt}\varepsilon \mathbf{E}(t) = \mathbf{M}^{-1}\mathbf{S}\mathbf{H}(t) - \mathbf{M}^{-1}\mathbf{F}(\vec{f}_E - \vec{f}_E^*) - \mathbf{J}(t),$$
$$\frac{d}{dt}\mu \mathbf{H}(t) = -\mathbf{M}^{-1}\mathbf{S}\mathbf{E}(t) + \mathbf{M}^{-1}\mathbf{F}(\vec{f}_H - \vec{f}_H^*),$$

where $\mathbf{E}$, $\mathbf{H}$ are vectors of nodal electric and magnetic field strength values, $\varepsilon$, $\mu$ are the electric permittivity and magnetic permeability, $\mathbf{S}$ describes a local differentiation matrix, $\mathbf{M}$ is a local DGFEM mass matrix and $\mathbf{F}(.)$ represents the flux expression coupling neighboring finite elements and $\mathbf{J}$ is the electric current density within the elements. The thin wire formulation following Holland [20] discretizes to

$$\frac{d}{dt}\mathbf{I}(t) = -v^2 \mathbf{D}_s \mathbf{q} + \mathbf{M}^{-1}\mathbf{F}(\vec{f}_I^* - \vec{f}_I) + \frac{\mathbf{E}_s}{L'},$$
$$\frac{d}{dt}\mathbf{q}(t) = -\mathbf{D}_s \mathbf{I} + \mathbf{M}^{-1}\mathbf{F}(\vec{f}_q^* - \vec{f}_q),$$

where $\mathbf{q}$ corresponds to a unit charge per wire length, v is the speed of interaction between charges, $L'$ denotes an inductance per unit length and $\mathbf{D}_s$ is the local derivative matrix along the wire.

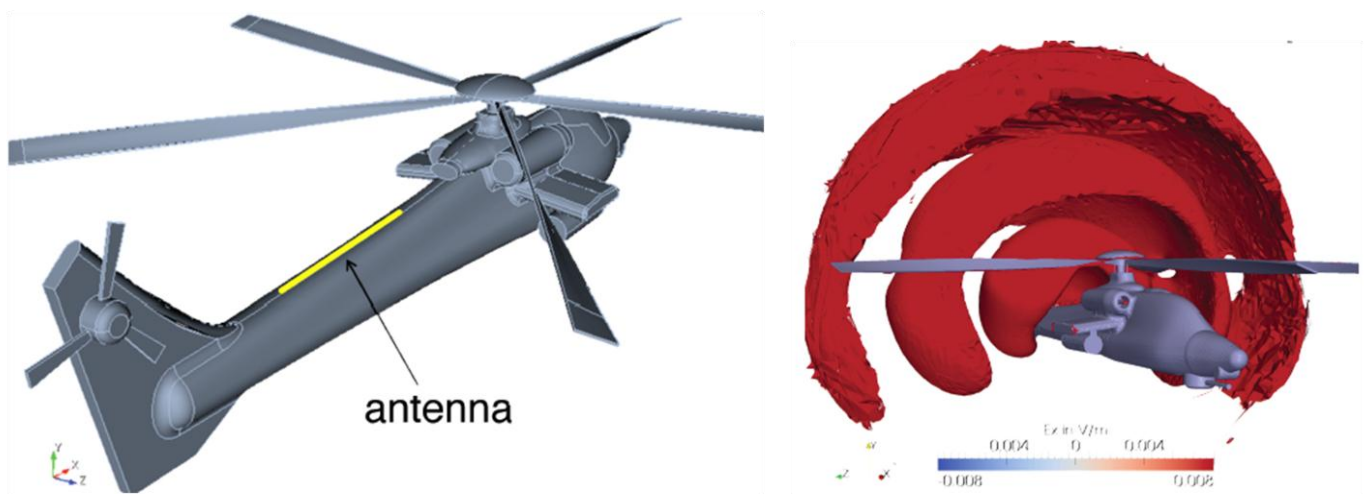

**Figure 6**: Field-cable coupled DGFEM simulation of a helicopter antenna (left) and the radiated electric field intensity [16]

The coupling involves the integration of the discrete electric field $\mathbf{E}_s$ inside each volume element (see Fig. 5). It adds the resulting electric voltage as an excitation source to the line model and the cable currents are used to calculate the right hand side excitation currents of the 3D field DGFEM formulation. The resulting high dimensional set of ordinary differential equations, i.e., the DGFEM-discretized 3D field equations and the 1D thin wire equation are solved simultaneously using an explicit time integration scheme such as e.g. an Low Storage Runge-Kutta method of $4^{th}$ order [15].

Fig. 6 shows the simulation result of wave propagation from a helicopter model featuring a thin wire antenna on its tail boom using this strongly coupled transient DGFEM field-cable formulation.

As the coupling of current MTL and 3D discrete electromagnetic field formulations is still subject to some severe restrictions concerning the validity of the MTL modeling assumptions for certain high-frequency EMC problems, this coupling is still matter of ongoing research efforts.

### C. Hybrid Discretization

Electromagnetic dosimetry simulations at radio frequencies using high resolution body phantoms may become a problem within FITD/FDTD simulations as soon as also large portions of an ambient environment have to be taken into account, possibly even including additional complex geometric details. Fig. 7 shows a simple example configuration of such an exposure situation with a body phantom positioned in some distance to an electromagnetic field source.

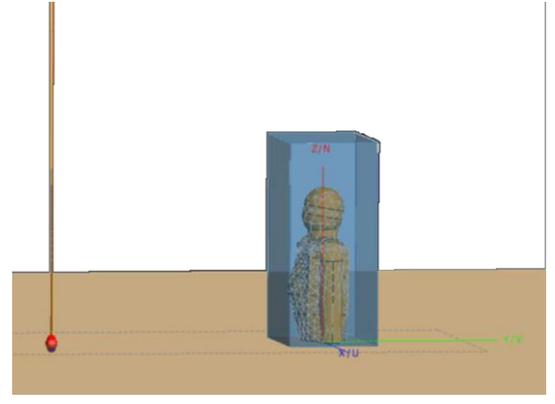

**Figure 7**: Exposure situation: Human body phantom on a perfectly conducting metal plane in some distance to an antenna [31]

Especially for such large scale exposure situations a weakly coupled method-of-moment (MoM) / FDTD multi-method approach can be adopted. First a MoM simulation is performed for the exposure situation, where the body model is replaced by either a homogeneous dielectric material body model or a perfectly electrically conducting (PEC) surface. Then this approximation of the field distribution between the body and the antenna is sampled on a closed Huygens surface around the body phantom. In a next step, a full three-dimensional FITD dosimetric SAR simulation is performed with a high-resolution voxel body phantom inside the "Huygens box" excited by the surface near field sources (see Fig. 8). While the weak MoM-FDTD coupled approach is an iterative approach, the iteration typically is stopped after the first step, since the FDTD calculated body currents are too small for a second MoM simulation and thus can be neglected.

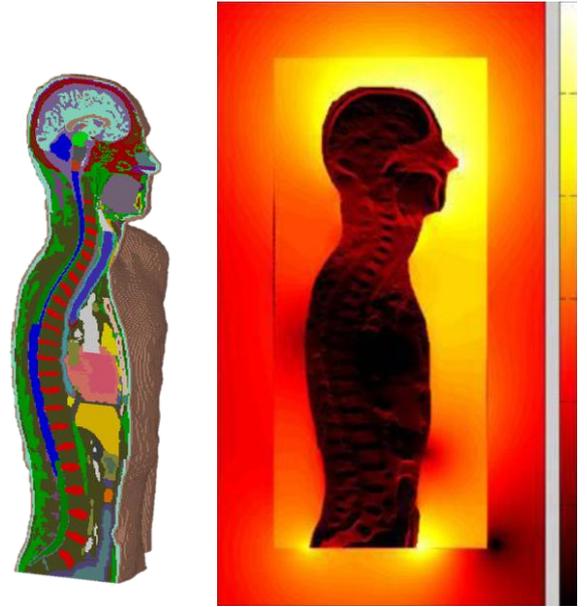

**Figure 8:** FDTD dosimetry simulation of a human body phantom excited by near surface source fields with the MoM calculated field values evaluated on the Huygens box surface [31]

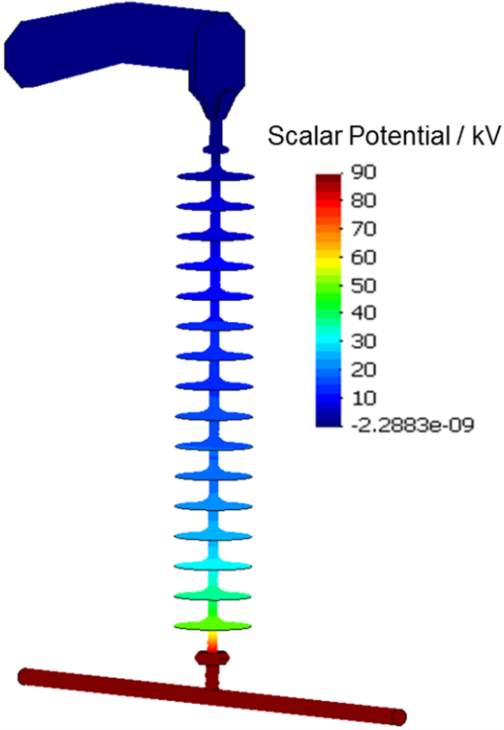

**Figure 9:** Electrostatic potential distribution on Lapp Insulators Rodurflex[TM] long rod insulator [28]

The optimization of nonlinear field stress grading in electric high-voltage system components such as surge arresters, microvaristor coated long rod insulators or microvaristor filled cable terminators requires nonlinear transient electro-quasistatic field simulations [14,28]. For these simulations a strongly coupled hybridization of a finite-element / boundary element method (FEM-BEM) has been proposed in [29]. With this formulation it is possible to model the attenuation behavior of the electric scalar potential, i.e., the BEM part models the "open" boundaries of the problem.

The strongly coupled FEM-BEM hybrid formulation in [29], here shown for electrostatic problems, yields the discrete formulation

$$\begin{pmatrix} \mathbf{A}_{ff} & \mathbf{A}_{fc} & 0 \\ \mathbf{A}_{cf} & \mathbf{A}_{cc}+\mathbf{D} & (-\tfrac{1}{2}\mathbf{M}^T+\mathbf{K}^T) \\ 0 & (-\tfrac{1}{2}\mathbf{M}+\mathbf{K}) & -\mathbf{V} \end{pmatrix} \begin{pmatrix} \mathbf{\Phi}_f \\ \mathbf{\Phi}_c \\ \mathbf{t} \end{pmatrix} = \begin{pmatrix} -\tilde{\mathbf{A}}_{fc}\mathbf{g}_D \\ -\tilde{\mathbf{D}}\mathbf{g}_D-\tilde{\mathbf{A}}_{cc}\mathbf{g}_D \\ (\tfrac{1}{2}\tilde{\mathbf{M}}-\tilde{\mathbf{K}})\mathbf{g}_D \end{pmatrix}$$

where the matrix blocks $\mathbf{A}_{ij}$ correspond to the FEM part of the formulation, the BEM-matrix block $\mathbf{M}$ corresponds to the discretization of the identity operator, $\mathbf{V}$ corresponds to the single-layer potential integral operator, $\mathbf{K}$ to the double layer-potential integral operator and $\mathbf{D}$ to the hypersingular integral operator. The BEM matrix blocks can be represented in a memory efficient way using matrix sparsification techniques as e.g. the Adaptive Cross Approximation [7]. The solution vector consists of $\mathbf{\Phi}_f$ inner node potential values, $\mathbf{\Phi}_c$ air-interface potential values and $\mathbf{t}$ the normal derivative values of the potential. The right hand side vector contains the predefined problem potential values as Dirichlet values $\mathbf{g}_D$. Fig. 9 shows a long rod insulator where the solution in the air part of problem is completely represented with the BEM formulation, thus avoiding the need to discretize the exterior air domain.

While the hybrid FEM-BEM coupled electro(-quasi-)static) formulation elegantly solves the problem of an "open" boundary, this comes at the prize of an additional complexity in the solution of the resulting algebraic systems of equations, where the challenge lies in an efficient preconditioning of the FEM-BEM-coupled system.

### D. Multi-Domain Coupled Formulation

Domain decomposition techniques are a widely established procedure in the numerical solution of boundary value problems, see e.g. [27],[32]. Quite often the segmentation of the domain into multiple domains blindly follows requirements of computational load balancing or algorithmic simplicity.

Depending on the problem to be solved, also more refined criteria for the decomposition of the computational domain into multiple subdomains may be useful. To highlight this, a nonlinear transient electro-quasistatic field formulation is considered. The FEM discretization [14,28] leads to a stiff nonlinear system of ordinary differential equations

$$\mathbf{M}_\varepsilon \frac{d}{dt}\mathbf{\Phi} + \mathbf{K}_\kappa(\mathbf{\Phi})\mathbf{\Phi} = \mathbf{b},$$

where $\mathbf{\Phi}$ is the vector of electric potential values, $\mathbf{M}_\varepsilon$ and $\mathbf{K}_\kappa$ denote the discrete div-grad operators w.r.t. permittivity and field dependent nonlinear conductivity. The Dirichlet boundary conditions are incorporated in the right hand side $\mathbf{b}$. The implicit time integration schemes used for these systems of stiff ordinary differential equations require long simulation times due to the large number of linear algebraic systems of high dimension that need to be solved repeatedly.

Using a multi-domain approach, the computational domain is split up into an interior subdomain corresponding to the dielectric materials including also the nonlinear conductive components and into an exterior subdomain corresponding to the air region with nodal vectors $\mathbf{\Phi}_1$ and $\mathbf{\Phi}_2$, respectively. As a result the discrete electro-quasistatic system is partitioned with

$$\begin{pmatrix} \mathbf{M}_{11} & \mathbf{M}_{12} \\ \mathbf{M}_{21} & \mathbf{M}_{22} \end{pmatrix} \frac{d}{dt}\begin{pmatrix} \mathbf{\Phi}_1 \\ \mathbf{\Phi}_2 \end{pmatrix} + \begin{pmatrix} \mathbf{K}_{11} & \mathbf{K}_{12} \\ \mathbf{K}_{21} & 0 \end{pmatrix}\begin{pmatrix} \mathbf{\Phi}_1 \\ \mathbf{\Phi}_2 \end{pmatrix} = \begin{pmatrix} \mathbf{b}_1 \\ \mathbf{b}_2 \end{pmatrix}.$$

Now, following a similar idea as in the coupled FEM-BEM approach [29] or the domain decomposition-based linear subspace information recycling approach published in [9], the linear exterior parts of the problem are now treated using proper orthogonal decomposition (POD) model order reduction (MOR) techniques aiming at a reduced dimension for the solution process.

Solving the full discrete system in time-domain allows saving snapshots of the solution. Then, a singular value decomposition of the snapshot matrix (see [12]), containing the system information, results in a low rank basis which represents the system dynamics and provides information for a projection operator. Constructing a projector only for the exterior nodes results in an $n \times p$ operator $\mathbf{P}$, where $n$ is the number of exterior domain degrees of freedom and $p$ the number of singular vectors used for the projection and $\mathbf{\Phi}_2 = \mathbf{P}\mathbf{z}_2$. Use of the projection operator $\mathbf{P}$ yields an ordinary system of equations

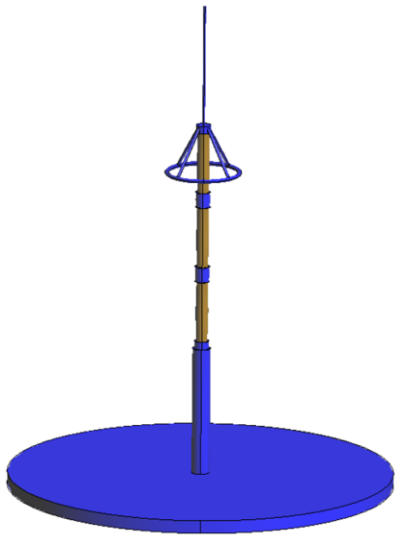
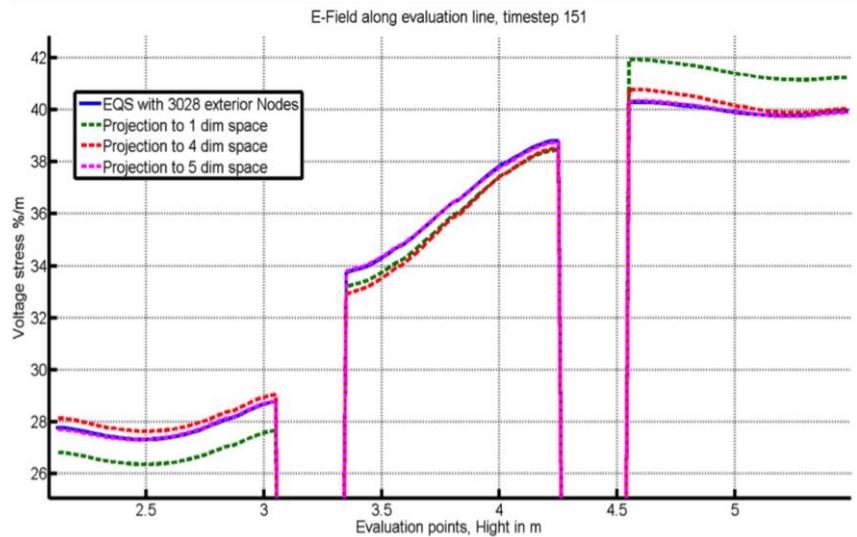

**Figure 10**: Electric field distribution along IEC surge arrester vertical axis: Results of the full nonlinear electro-quasistatic simulation and the three MOR projected simulations with $p$=1,4,5

$$\begin{pmatrix} \mathbf{M}_{11} & \mathbf{M}_{12}\mathbf{P} \\ \mathbf{P}^\top \mathbf{M}_{21} & \mathbf{P}^\top \mathbf{M}_{22}\mathbf{P} \end{pmatrix} \frac{d}{dt} \begin{pmatrix} \Phi_1 \\ \mathbf{z}_2 \end{pmatrix} + \begin{pmatrix} \mathbf{K}_{11} & \mathbf{K}_{12}\mathbf{P} \\ \mathbf{P}^\top \mathbf{K}_{21} & \mathbf{0} \end{pmatrix} \begin{pmatrix} \Phi_1 \\ \mathbf{z}_2 \end{pmatrix} = \begin{pmatrix} \mathbf{b}_1 \\ \mathbf{P}^\top \mathbf{b}_2 \end{pmatrix}.$$

Here, the number of degrees of freedom in the exterior air domain is reduced from $n$ to dimension $p$.

The validity and effectiveness of this multi-domain coupled approach is shown in a nonlinear electro-quasistatic 2D field simulation of an IEC norm surge arrester design [28,30]. Fig. 10 shows simulated electric field distribution results along the vertical surge arrester axis, where the results of the MOR projected simulations with $p$=1,4,5 are compared to a full simulation with 3,900 degrees of freedom with an exterior domain problem dimension $n$=3,028 (about 78% of the whole problem size). The MOR solution for $p$=5 coincides with the full solution, which corresponds to a dimension reduction factor of about 600. This reduction in the problem dimension allows then for a faster optimization of the nonlinear material characteristics.

## V. CONCLUSION AND OUTLOOK

Many high and low frequency electromagnetic field simulation methods have reached a high level of maturity. Today's challenges are often due to multiscale or multiphysics phenomena. In this paper a classification of coupled problems was presented with corresponding examples of coupled formulation, e.g., for scale bridging (cable models, hybrid discretizations, MOR techniques) and multiphysical simulation (in particular the important coupling with the heat equation).

The convergence issue of coupled time-domain simulations was discussed and an iterative algorithm to increase stability was proposed. One important result that applies also to co-simulations without iterations is that reordering the computational sequence of the subsystems can improve stability. This can be mathematically verified by analyzing the Jacobians.

When looking into the future of coupled simulations, the data exchange will be an increasingly important factor. Today, most implementations for coupled simulation are still single-threaded or exchange their complete volumetric data at synchronization steps, e.g. one core computes the electromagnetic field and another one the thermal distribution. This procedure may become increasingly expensive in terms of time and energy consumption while the costs per flops will further reduce in the future. According to SciDAC reports, [8], medium-sized clusters will have in the range of 400.000 cores by 2020 such that data-movement efficient parallelization is an important issue for large-scale computations.

Possibly the traditional way of splitting problems by physics must be replaced by a splitting in space, e.g., the computational domain is decomposed into subdomains and each core computes electromagnetic and heat effects for its subdomain locally, which then will only require the exchange of boundary information.

### ACKNOWLEDGEMENTS

The authors gratefully acknowledge the support of the BMBF grant 03MS648E "SOFA: Coupled Simulation and Robust Optimization in Virtual Vehicle Design" and the collaboration with Delphi Deutschland GmbH, Wuppertal, Lapp Insulators GmbH, Wunsiedel, and the Computersimulationstechnik AG, Darmstadt.

AUTHORS NAME AND AFFILIATION

Markus Clemens[1], Sebastian Schöps[2], Carsten Cimala[1], Nico Gödel[1], Simon Runke[1], Daniel Schmidthäusler[1], Thomas Timm[1]

[1]Bergische Universität Wuppertal,
Chair of Electromagnetic Theory,
Rainer-Gruenter-Str. 21,
42119 Wuppertal, Germany
clemens@uni-wuppertal.de
www.tet.uni-wuppertal.de

[2]Technische Universität Darmstadt, Graduate School CE/TEMF,
Dolivostraße 15,
64293 Darmstadt, Germany
schoeps@gsc.tu-darmstadt.de
www.graduate-school-ce.de